\documentclass[preprint,showpacs]{revtex4-1}
\usepackage{enumerate}
\usepackage{graphicx}
\usepackage{natbib}
\usepackage{amsmath}
\usepackage{time}
\usepackage{subfig}
\usepackage{setspace}
\usepackage{epstopdf}
\input epsf.sty
\begin{document}

\title{Experimental Signatures of the Quantum-Classical Transition
in a Nanomechanical Oscillator Modeled as a Damped Driven Double-Well
Problem}
\author{Qi Li, Arie Kapulkin, Dustin Anderson, Shao Min Tan, 
Arjendu K. Pattanayak}
\affiliation{
Carleton College, 1 College Street, Northfield, Minnesota, 55057
}
%\address{%
%\begin{minipage}[t]{6.0in}
\begin{abstract}
We demonstrate robust and reliable signatures for the transition 
from quantum to classical behavior in the position probability
distribution of a damped double-well system using the Qunatum State Diffusion
approach to open quantum systems.  We argue that these signatures are 
within experimental reach, for example in a doubly-clamped nanomechanical beam.
\typeout{polish abstract}
\end{abstract}
%\end{minipage}

\maketitle
%\narrowtext

%\doublespacing

%\linespread{2.0}

\section{Introduction}

The difference between quantum and classical behavior is of fundamental
interest with many experimental consequences in contexts such as 
superconducting quantum interference devices (SQUIDs), cold atom
systems, and significantly in nanoelectromechanical systems (NEMS).
NEMS are small, oscillate at high frequencies and are typically maintained
at low temperatures. Most of their parameters are adjustable, making them 
an ideal setting to explore the transition from quantum to classical behavior.
Recent experimental results\cite{Chan2011,Teufel2011}, indicate that such 
systems can now be prepared and observed in their ground state, thus enabling 
the study of genuinely non-classical behavior in the dynamics. The exciting 
prospect of being able to observe the quantum-classical transition in detail 
has contributed to interest in quantum mechanical nanoelectromechanical systems
(occasionally refered to as QEMS in the literature), whose study is an active and 
rapidly expanding field of research\cite{Poot2012273}.

A seminal theoretical study of the quantum--classical transition in 
NEMS was {carried out} by Katz {et
al}\cite{lifshitz,*PhysRevLett.99.040404} {using} a nonlinear 
resonator, {which is} a paradigmatic model for several common types of 
NEMS that yields quantum behavior distinct from the classical. They 
considered an isolated resonator as well as one modeled as an open 
system (coupled to the environment) in search of experimental signatures 
of such quantum behavior.  Specifically, they examined the dynamics of
the classical phase space distribution in such a system compared to that
of the quantum-mechanical Wigner function, and looked at a zero temperature
and a finite temperature environment. They were able to find differences in 
experimentally accessible signatures: In particular, the quantum 
mechanical version of the resonator has non-zero probability of being 
found in a position where the classical resonator has zero probability 
of being found. However, these differences are relatively small and blur 
away quickly when experimental noise or finite temperature effects
through thermal noise are considered. 

In this paper, by considering a slightly different but equally accessible 
nonlinear oscillator, we are able to obtain more pronounced experimental 
signatures of the quantum-classical transition; these are arguably more
visible at finite temperatures and in the presence of experimental 
noise. Further, we show that rather than comparing just quantum and 
classical behavior, there is insight to be gained from studying the 
continuous transition where the classical behavior emerges as the 
limiting case of the quantum dynamics. That is, although systems are 
fundamentally described by quantum mechanics, their behavior changes 
as we increase the size of the system (affecting the characteristic 
action scaled by $\hbar$), the effect of the environment (decoherence) 
or other parameters of the system. For our system, we indeed recover 
the classical behavior as the limiting case. The details of the 
emergence of the classical limit are illuminating, and in particular, 
the signatures do not always change monotonically, thus establishing
that comparing a single quantum solution with a classical solution leads 
to an incomplete picture. 

Katz {\em et al} considered the sytem with the Hamiltonian 
\begin{equation}
H_{sys}=\frac{1}{2}p^2+\frac{1}{2}x^2+\frac{1}{4}x^4-xF\cos\omega t.
\label{duffingEOM}
\end{equation}
This quadratic nonlinear oscillator is within experimental reach in the
near-quantum regime, and is an excellent system to study.
However, even more interesting effects such as quantum tunneling 
obtain if we change the sign so that the $x^2$ term is
negative\cite{carr2001}.  This
yields the so-called double-well Duffing oscillator, which has the 
Hamiltonian 
$H_{sys}=\frac{1}{2}p^2-\frac{1}{2}x^2+\frac{1}{4}x^4-xF\cos\omega t$,
and allows for clearer signatures of the classical to quantum
transition as we demonstrate below.

This paper is organized as follows: We start by motivating the modeling 
of a specific NEMS as a doubly-clamped beam to show the connection
between experimental paramters and model parameters. 
We then sketch the quantum state diffusion (QSD) approach, where Lindblad 
operators act within stochastic Schrodinger equations to incorporate the 
effect of the environment, used to model the behavior of such an 
oscillator understood as an open quantum system. 
These two allow us to establish that changing the sign of the quadratic 
term is within current experimental capabilities and that such experiments 
are in or close to the realm where the quantum to classical transition can 
be explored. We then present results and conclude with our analysis.

\section{NEMS modeled as a doubly-clamped beam}
Many nanoelectromechanical devices can be modeled as doubly-clamped
beams driven near resonance and continuum mechanics continues to 
serve as an adequate model even at the submicron 
scale\cite{number7,number9,number4}.
Usually magnetomotive\cite{Cleland1999256} or optical\cite{number7}
actuation is used to study the resonance behaviors of NEMS such as
dynamically induced bistability, hysteresis and effects of parameter
attenuation on energy dissipation.
The resonant behavior of the doubly-clamped NEMS structure has been
experimentally identified with the fundamental bending mode\cite{number3}
of the problem. We sketch here the theory allowing us to relate the resonance 
frequency to characteristic parameters of NEMS and to determine the 
scale of the system with respect to $\hbar$, establishing the regime of 
the transition to classical behavior in these systems.

Starting with the assumption of an ideal beam of rectangular
cross-sectional area, we can obtain the potential energy $V$
of a doubly-clamped beam under tension along the longitudinal
direction. This consists of two parts\cite{modeling}: elastic and
bending potential energy,
\begin{equation}
V_{elastic}= T(\int_{0}^{l_0}\sqrt{1+(y')^2}dx-l_0)
\label{elastic energy}
\end{equation}
and
\begin{equation}
V_{bending}= \frac{EI}{2}\int_{0}^{l_0}(y^{''})^2 dx.
\end{equation}
$T$ is the external force on the beam in the form of the tension
in the longitudinal direction along the beam, $E$ is the elastic
modulus, $I$ is moment of inertia of the cross-section and $l_0$
is separation of two ends that clamp the beam. 
Note that $l_0$
need not be the length of the beam when it is neither stretched
nor compressed because a static doubly-clamped beam may already be
buckled due to inherent tension along the beam. 
$V_{elastic}$ is due to tension built up in the beam, which causes the
elastic beam to deform and therefore gain energy. This is usually
referred to as strain energy\cite{elasticity} while $V_{bending}$
is purely due to bending.
The quantity $EI$, termed ``flexural rigidity'', denotes the force
necessary to bend a beam by a unit of curvature.

The tension $T$ along the beam has two parts\cite{lifshitzReview}:
firstly, inherent tension $T_0$ (either from compression at both ends
of the beam or manufacturing process) and additional tension
$\Delta T$ (due to transverse motion that stretches the beam),
whence $T=T_0+\Delta T$. When the beam is stretched, its length changes
and therefore the additional tension $\Delta T$ is given by fractional
change in length multiplied by elastic modulus and cross section
area $A$\cite{lifshitzReview}:$\Delta T = \frac{l-l_0}{l_0}E A$,
where $l$ refers to the actual length of the buckled beam. We can
rewrite this length as
\begin{equation}
l = l_0+\Delta l = \int_0^{l_0}dx \sqrt{1+(y')^2}\approx
\int_0^{l_0}dx(1+\frac{1}{2}(y')^2) = l_0+\frac{1}{2}\int_0^{l_0}(y')^2 dx.
\end{equation}
This yields that the change in tension 
$\Delta T=\frac{E A}{2l_0}\int_0^{l_0}(y')^2 dx$ whence the total
tension 
\begin{equation}
T = T_0+\frac{E A}{2l_0}\int_0^{l_0}(y')^2 dx. 
\end{equation}
Substituting these expressions for $\Delta l$ and $T$ in the expression
for elastic potential energy above, we get 
\begin{equation}
V_{elastic} = 
 \frac{T_0}{2}\int_0^{l_0}(y')^2 dx+\frac{E A}{4l_0}[\int_0^{l_0}(y')^2 dx]^2.
\label{elastic energy1}
\end{equation}
Thus, the total potential energy is
\begin{equation}
\begin{array}{rcl}
V &=& V_{elastic} + V_{bending}\\
  &=& \frac{T_0}{2}\int_{0}^{l_0}(y')^2 dx + \frac{E A}{4 l_0}[\int_0^{l_0}(y')^2 dx]^2 + \frac{E I}{2}\int_0^{l_0}(y'')^2 dx.
\end{array}
\label{potential}
\end{equation}
At low temperatures, as is often true for NEMS experiments, it is
sufficient to consider only the fundamental mode. 
Mode amplitude solutions for this fundamental mode $y(x)$ are of 
the form \cite{carr2002}
$y(x)=\frac{1}{2}Y[1-\cos(2\pi\frac{x}{l_0})]$, where
$Y$ is the transverse displacement of the center point of the beam.
When we make this substitution, the fundamental mode amplitude 
$Y$ -- that is, the location of the vibrating central point -- 
is effectively the position variable of a single particle evolving 
in a potential $V$, which as a function of $Y$ is given by
\begin{equation}
\begin{array}{rcl}
V(Y) &=& \frac{E I \pi^4}{l_0^3}Y^2 + \frac{\pi^2 T_0}{4 l_0}Y^2 + \frac{E A \pi^4}{16 l_0^3} Y^4\\
 &=& Y^2(\frac{E I \pi^4}{l_0^3}+\frac{\pi^2 T_0}{4 l_0}) + Y^4 \frac{E A \pi^4}{16 l_0^3}.
\end{array}
\label{potential1}
\end{equation}
Here $I$, the moment of inertia of rectangular cross section is
$\frac{ab^3}{12}$\cite{elasticity1}, where $a$ is the thickness of the beam
and $b$ is its width. $I$ varies with the geometry of the beam for 
specific experimental setups. Further, depending on whether inherent 
tension $T_0$ compresses or stretches the beam, the sign of $T_0$ can 
be negative or positive.  
Compressive inherent tension along the beam can arise from manufacturing 
processes, and can vary from $200$ MPa to $1$ GPa\cite{carr2002}. In 
general, the inherent tension is not known a priori in an experimental 
setup. Instead, by considering the continuum mechanics model and fitting 
the experimentally measurable resonance frequency, the inherent tension 
can be calculated\cite{number3}. For sufficiently large inherent 
compressive tension, $T_0<-\frac{4\pi^2EI}{l_0^2}$, we obtain a 
double-well potential and the beam is in a state of ``Euler 
instability''\cite{number4,number6}, meaning that it is buckled. 

Thus a single-well potential or a double-well potential can be 
experimentally achieved in doubly-clamped NEMS structures and these
systems can then be studied for their dyamical behavior.
The dynamics of a single-well NEMS actuated near the fundamental bending 
mode frequency, has been extensively studied\cite{basinOfAttraction,number5-peanoAndThorwart}
but that of the double-well case has not been yet been fully explored. 
The results of Ref.~\cite{number3} suggest that it within
experimental reach and the above clarifies that in general tuning the 
characteristic parameters should enable switching from a single-well to 
a double-well potential.

\section{Open Quantum Systems}
We now turn to the quantum version of the model for the problem and
connect characteristic parameters with the degree of `quantumness' of
the system.
An open quantum system with weak system-environment coupling and a
Markovian environment is modeled by a master equation
\begin{equation}
\begin{split}
\dot{\hat{\rho}}(t)=& \frac{-i}{\hbar} \left[\hat{H},\hat{\rho}\right]-
\frac{1}{2}\sum_j \left(\hat{L_j}^\dagger\hat{L_j}\hat{\rho}+
\hat{\rho}\hat{L_j}^\dagger\hat{L_j}\right)\\
 & + \sum_j \hat{L_j}\hat{\rho}\hat{L_j}^\dagger,
\label{masterEq}
\end{split}
\end{equation}
where $\hat{L_j}$ is the Lindblad operator representing the system-environment
interaction and $\hat{\rho}$ is the reduced density operator.
The quantum state diffusion (QSD) approach corresponds to solving
quantum trajectories which are unravellings of the master equation. This
allows for numerical efficiencies, and physical insights not
available via the master equation.  Specifically, we can use the
QSD numerical library\cite{qsd} to solve a stochastic version of the
Schr\"odinger equation
 \begin{equation}
 \begin{split}
 d|\psi\rangle=&
 \frac{i}{\hbar}\hat{H}|\psi\rangle dt\\
 & + \sum_j \left(\langle \hat{L_j}^{\dagger}\rangle \hat{L_j} -\frac{1}{2}\hat{L_j}^{\dagger}\hat{L_j}
-\frac{1}{2}\langle \hat{L_j}^{\dagger}\rangle\langle\hat{L_j}\rangle \right)|\psi\rangle dt\\
 & +\sum_j \left( \hat{L_j}-\langle \hat{L_j} \rangle \right) |\psi \rangle d\xi_j.
\label{sde}
\end{split}
\end{equation}
The solution $|\psi(t)\rangle$ to Eq.~(\ref{sde}) is 
a ``quantum trajectory'', and we obtain 
$\hat{\rho}=\frac{1}{M}\sum^M_i|\psi^i\rangle\langle\psi^i|$, as the mean
over an ensemble of $M$ normalized pure state projection
operators\cite{qsd}. This $\hat{\rho}$ satisfies the Lindblad
master equation Eq.~(\ref{masterEq}). That is, starting with an
ensemble of identical pure states, we obtain the time evolution of
multiple trajectories that evolve into an ensemble of different pure
states due to interaction with the environment; this is the density 
matrix.

For the open quantum double-well Duffing oscillator
$\hat{H}$ and $\hat{L_j}$ in Eq.~(\ref{sde})
are chosen as\cite{akap,brun,ref7inAKAP,lifshitz,*PhysRevLett.99.040404}
\begin{eqnarray}
&& \hat{H}=\hat{H_D}+\hat{H_R}+\hat{H_{ex}} \\
&& \hat{H_D}=\frac{1}{2m}\hat{p}^2 + \frac{m\omega_0^2}{4l^2}x^4-\frac{m\omega_0^2}{2}x^2, \label{duffinghamiltonian} \\
&& \hat{H_R}=\frac{\gamma}{2}(\hat{x}\hat{p}+\hat{p}\hat{x}) \\
&& \hat{H_{ex}}=-gml\omega_0^2\hat{x}\cos(\omega t) \\
&& \hat{L_1}=\sqrt{\gamma\left(1+\bar{n}\right)}\left(\sqrt{\frac{m\omega_0}{\hbar}}\hat{x}+i\sqrt{\frac{\gamma}{m\omega_0 \hbar}}\hat{p}\right), \label{l1} \\
&& \hat{L_2}=\sqrt{\gamma\bar{n}}\left(\sqrt{\frac{m\omega_0}{\hbar}}\hat{x}-i\sqrt{\frac{\gamma}{m\omega_0 \hbar}}\hat{p}\right). \label{l2}
\end{eqnarray}
where $\gamma$ is the strength of coupling 
between the oscillator and thermal bath environment. 
We now define $\hat{x}$ and $\hat{p}$ in unitless forms
$\hat{Q}=\sqrt{m\omega_0/\hbar}\hat{x}$ and
$\hat{P}=\sqrt{1/m\omega_0\hbar}\hat{p}$
respectively\cite{brun,ref7inAKAP} yielding the dimensionless set of 
equations
\begin{eqnarray}
&& \hat{H^{'}_{D}}=\frac{1}{2}\hat{P}^2+\frac{\beta^2}{4}\hat{Q}^4-\frac{1}{2}\hat{Q}^2 \label{duffingH} \\
&& \hat{H^{'}_R}=\frac{\Gamma}{2}\left(\hat{Q}\hat{P}+\hat{P}\hat{Q}\right) \\
&& \hat{H^{'}_{ex}}=-\frac{g}{\beta}\hat{Q}\cos(\Omega t) \\
&&
\hat{L^{'}_1}=\sqrt{\Gamma(1+\bar{n})}\left(\hat{Q}+i\hat{P}\right)
\label{l1p} \\
&&
\hat{L^{'}_2}=\sqrt{\Gamma\bar{n}}\left(\hat{Q}-i\hat{P}\right)\label{l2p} 
\end{eqnarray}
where the time in measured in units of $\omega_0^{-1}$ so that $\Gamma =
\gamma/\omega_0, \Omega = \omega/\omega_0$. 
In Eqs.~(\ref{l1}, \ref{l2},\ref{l1p},\ref{l2p}),
$\bar{n}=\left(e^{\hbar\Omega/k_BT} - 1\right)^{-1}$ is the Bose-Einstein
distribution of thermal photons representing the effect of the environment 
at finite $T$ and is evaluated at the natural frequency of the 
oscillator\cite{quantumnoise}. This dimensionless
formulation isolates the scaling factor $\beta=\sqrt{\frac{\hbar}{S}}$ as the
single length scale of the problem, where $S$ is the classical action,
such that $\beta^2 = \frac{\hbar}{ml^2\omega_0}$.
For the classical equation of motion for a damped Duffing oscillator,
which is $\frac{d^2x}{dt^2}+2\Gamma \frac{dx}{dt} +
\beta^2x^3-x=\frac{g}{\beta}\cos(\Omega t)$,
solutions to the classical equation of motion are unchanged with respect
to different values of $\beta$ except for the length scale of the
phase space. However, changing $\beta$ changes the quantum solution
considerably.  

Comparing the quartic potential Eq.~(\ref{potential1}) for the motion of 
the fundamental mode of the nanomechanical resonator with the potential
term in the Duffing Hamiltonian Eq.~(\ref{duffinghamiltonian}), it is
straightforward to see that for the quantum nanoelectromechanical resonator 
\begin{equation}
\beta^2 \equiv \frac{\hbar}{S}  
= \frac{\hbar l_0}{8\pi^2} \sqrt{\frac{A}{2E\rho{I}^3[(-1)(1 + \frac{T_0l_0^2}{4\pi^2EI})]^3}}
\label{beta}
\end{equation} 
where, as noted in section II, we restrict our consideration to 
resonators of the double well shape with $T_0<-\frac{4\pi^2EI}{l_0^2}$.
The scaling parameter $\beta^2$ determines the way in which the action
of the nanomechanical system scales relative to $\hbar$. 

As such, studying the system as $\beta$ is changed allows us to explore 
the quantum to classical transition. Previous theoretical 
work\cite{akap,brun,ref7inAKAP} for the Duffing problem focused on the
behavior of the quantum trajectories has confirmed that $\beta \rightarrow 0$ 
indeed recovers the classical limiting behavior, including the presence
of 'strange attractors' of chaos in the phase space, while $\beta=1$ 
shows behavior distinctive of the quantum regime. There is interesting 
physics in the intermediate regime\cite{akap}, and in particular the 
details of the transition as $\beta$ changes are informative (as below). 
As shown in Table (\ref{table1}), doubly-clamped nanoelectromechanical 
beams can be made of different materials including Si\cite{carr2002,carr2001}, 
single-walled carbon nanotubes (SWNT) or multi-walled (MWNT) 
nanotubes\cite{number7} and even metals like gold\cite{goldbeam}. 
The analysis above for the Duffing oscillator also works for a doubly-clamped 
Pt nanowire\cite{ref10inNumber7}, which has a high fundamental frequency
(greater than 1 GHz). The characteristic parameters given in
Table (\ref{table1}) for example show that the current experimental 
setups can be tuned within the range of $\beta$ considered in this paper.
Specifically, for the SWNT experiment of Witkamp et al\cite{number3},
(see Table \ref{table1}), they used a device of radius $r=1.6 nm$
and length $l_0=1.15 \mu m$. This yields, using their other device
parameters and an $I = \frac{\pi r^4}{4}$ that 
\begin{equation}
\beta^2 = 5.34 X 10^{-6} X
(\frac{\lambda}{4\pi^2}-1)^{-3/2}
\end{equation}
where $\lambda = \frac{T_0l_0^2}{EI}$ is greater than $4\pi^2$ 
to yield the double-well shape. If $\lambda$ is in the range 
$(39.5 - 60)$, we get $\beta$ in the range $(0.65 - 0.04)$, which as we 
see below, is precisely the range needed to map the transition. 
Witkamp et al report $\lambda = 26$, so this is clearly
well within reach of current experiments.

\begin{table}[htbp]
\caption{Four common types of doubly-clamped structure of NEMS that
have been realized experimentally.  The characteristic parameters can be
used to find the fundamental mode frequency.}
\scalebox{0.6}{
    \begin{tabular}{ | l | l | l | l | l |}
    \hline
    Type & Length, $l_0$ (nm) & Diameter (nm) & Density $\rho$ (kg/m$^3$) & Elastic Modulus$E$(TPa)\\ \hline
    Si\cite{carr2001} & 50-10$^4$ & 5$\times$10 to 10$\times$20 & 2330 & 0.137\\ \hline
    SWNT\cite{number3},\cite{number7} & 50-10$^4$ & 1~ & 1930 & 1.25\\ \hline
    MWNT\cite{number7} & 50-10$^4$ & 20 & 1930 & 1.25\\ \hline
    Pt nanowire\cite{ref10inNumber7} & 1300 & 43 & 168 & 21090\\
    \hline
    \end{tabular}}
\label{table1}
\end{table}

There are other parameters of interest related to the system-environment
interactions. In particular, the experimentally measurable quality factor $Q$
quantifies energy lost in comparison to total energy of the resonator during
one complete driving period which can be related to the damping $\Gamma$. 
A comprehensive account for the mechanisms of quality factor, $Q$, in NEMS has
not been established\cite{goldbeam}. but for the case of doubly-clamped beam,
$Q$ can arise from clamping, thermal elastic damping, as well as defects of
crystalline structure of the material\cite{number9}. Usually, $Q$ can be
measured experimentally using magnetomotive
techniques\cite{Cleland1999256}.
For low temperatures (20 K and below) $Q$ can be as large as of the order 
of $10^3$\cite{basinOfAttraction}.  Lowering the temperature further to 
the milikelvin range further depresses energy dissipation and
recent experiments\cite{goldbeam,number3_2009} have achieved quality
factors on the order of $10^5$. Although these experiments have pushed
towards high $Q$ and correspondingly low $\Gamma$, it is clearly
possible to make systems more dissipative, thus allowing the exploration
of a large range in $\Gamma$.

\section{Results and Discussions}
The solutions to Eq.~(\ref{sde}) yield ``quantum trajectories'' which have
been previously discussed in detail in the context of the recovery of
classical behavior from quantum mechanics\cite{brun,akap,ref7inAKAP}. What
we focus our attention on below is ensemble-and-time-averaged
asymptotic behavior of the system, defined through the probability
distributions
\begin{equation}
P_{avg}(x)=
\sum_{i=1}^{M}\sum_{k=1}^{N}\frac{1}{MN}\langle\psi^i_k(t)|\psi^i_k(t)\rangle.
\end{equation}
with $t_{\rm transient} + 4\pi > t> t_{\rm transient}$. In doing this we 
are averaging over multiple $M$ trajectories generated in QSD simulations 
to yield the behavior of density matrices accessible the laboratory.
We also eliminate phase-dependent idiosyncracies of these distributions 
{\emph and correlate better with experimental capabilities regarding the 
measurement of the dynamics of the quadratures by}
time-averaging trajectories $|\psi^i_k(t)\rangle$ with $N$ samples 
of each trajectory over two driving periods, taken after a suitably long 
transient time $t_{\rm transient}$. 

The time averaging is not necessary but is useful because (a) in exploring 
the quantum to classical transition, the properties that signal change most 
clearly as parameters are varied are asymptotic, global characteristics of 
the system rather than unrepresentative ``snap shots'' at specific times.  
Also (b), the time-averaging eliminates experimental difficulties in determining 
the phase of the driving relative to the observation, as well as allowing 
one to quickly build statistics.

\begin{figure}[htbp]
\includegraphics[scale=0.5]{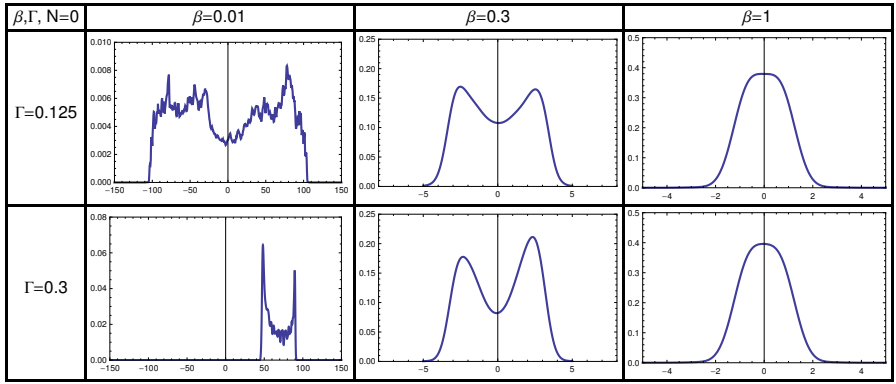}
\caption{$P_{avg}(x)$ plots for various parameters. The degree of
quantumness $\beta$ increases from left to right as $0.01, 0.3, 1.0$. 
The values for $\Gamma = 0.125, 0.3$.}
\label{grid1}
\end{figure}

We have explored $P_{avg}(x)$ at both zero temperature (that is,
$\bar{n}=0$) and finite temperature for several $\beta$; we report here
$\beta= 0.01, 0.3$ and $1.0$. For each of these values, as we change the
coupling strength $\Gamma$, we also looked for signatures of the quantum 
to classical transition through the changes in $P_{avg}(x)$. 
Fig.~(\ref{grid1}) shows $P_{avg}(x)$ at zero temperature. With high
dissipation ($\Gamma=0.3$) we see what we term nonmonotonicity in the
behavior of the probability distribution as a function of $\beta$.
Specifically, the probability distribution changes as follows: (a) At
$\beta = 0.01$ it is an asymmetric object centered in a single well, 
followed by (b) at $\beta = 0.3$ an essentially symmetric double-peak 
structure across both wells and then (c) at $\beta = 0.1$ a symmetric 
single peak straddling the central potential barrier. 

We understand these results as follows:
For $\beta=0.01$, the most classical case, the dissipation results
in the particle being unable to overcome the central potential barrier
even with driving and therefore being confined in a single well. The
probability peaks at both ends of the probability distribution correspond
to the classical turning points of the oscillator orbit in phase space.
That is, the dynamics in this situation are identical to those that
would be generated by a classical trajectory, and unsurprisingly this
means that the probability distribution is also entirely explained by
classical considerations.

As we increase the quantum-ness of the system (decrease the scale of the
system) to $\beta=0.3$,
the particle acquires a nonzero probability of being in both wells.
This is classically impossible given that it is a zero-temperature
problem with high damping. The inter-well transitions could indeed
occur because, as the system gets more quantum-mechanical with
$\beta$ increasing, the scale of the Lindblad term effectively also
increases. It is therefore possible that the particle continues to act
essentially classically but the increased effect of the noise leads to a
classical noise-activated hopping over the barrier. However, our studies
show that the particle is behaving entirely quantum-mechanically
during the interwell transition, whence this transition is a quantum 
tunneling effect.
Specifically, we have seen that the Wigner function for the
particle shows classically forbidden negative regions throughout its
dynamics. 
We have also seen that the energy expectation value remains 
negative while the expectation value of the position transits past the 
origin\cite{akap}. 
Finally, we have found instances in time when a wave-packet has 
significant probability on either side of the barrier, which is 
incomensurate with it being a localized wavepacket behaving like a 
classical particle hopping over the barrier. 
Incidentally, the last two demonstrate one advantage of using 
the QSD wavefunction-based approach, since such information is unavailable 
in the master equation approach. 
The symmetric double-peak structure of $P_{avg}(x)$ at this value of
$\beta$ indicates that the ensemble average over the wave packet behavior 
leads to an asymptotically equal probability in either well.

For $\beta=1$, the most quantum case, the probability is peaked at the
origin. There are multiple ways of understanding this. The first is that the
zero point energy is such that the lowest states of the potential are
above the barrier. As such the probability is peaked in the middle of the
quartic larger well, ignoring the quadratic bump at the bottom. Another way
of understanding this situation is to realize that as $\beta$ is increased
above $0.3$, the scale of the system changes such that the two peaks of the
symmetric distribution from $\beta =0.3$ come closer together until they
merge.

These three figures show not only that there is a clearer difference 
between the quantum and the classical probability behavior in the 
double-well Duffing oscillator compared to the single-well version, 
there are interesting things to be learned from studying the intermediate 
regime.

As as alternative case, we also present results from slightly lower
dissipation, with $\Gamma = 0.125$. In this case, even at the classical 
limit, the dissipation is not strong enough to confine the particle in 
one well. Specifically, the particle trajectory is that of a strange 
attractor across both wells\cite{akap,brun,ref7inAKAP} which yields the 
unusual shape for the probability distribution. The difference in the
classical behavior does not persist, and the distributions at higher 
$\beta$ for this $\Gamma$ value follow the same pattern as those for 
$\Gamma =0.3$. As such, comparing the two sets of figures, the 
results at $\Gamma =0.3$ show a more dramatic transition from the
classical to the quantum regime. 

\begin{figure}[htbp]
\includegraphics[scale=0.5]{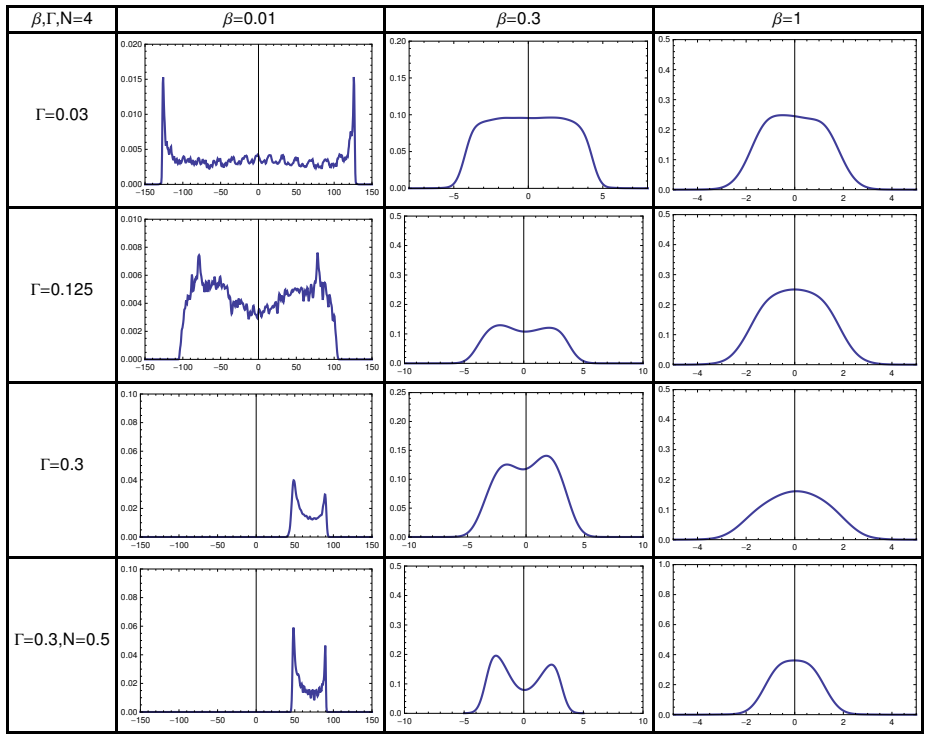}
\caption{$P_{avg}(x)$ plots for various parameters at finite temperature. 
For the first three rows, the number of thermal photons 
$\bar{n}=4$ and $\beta$ increases 
from left to right as $0.01, 0.3, 1.0$. $\Gamma$ increases down the grid 
as $0.03, 0.125, 0.3$. The last row has $\Gamma=0.3$ and $\bar{n}=0.5$
and shows a smooth transition from the zero temperature case in 
Fig.~(\ref{grid1}).}
\label{grid2}
\end{figure}

At even lower values of $\Gamma$, the vanishing dissipation means that 
the wavefunctions do not localize in a basis as happens for greater
dissipation\cite{qsd}. This leads to the standard semiclassical
convergence problem for the calculations of $P_{avg}(x)$. 
However we are able to compute $P_{avg}(x)$ at finite temperature
for each set of $\beta$ and $\Gamma$ and these are presented in 
Fig.~(\ref{grid2}). We see quickly that the results are similar to those at 
zero temperature in Fig.~(\ref{grid1}) though we can see that
temperatures corresponding to $\simeq 4$ thermal photons begins to 
eliminate the structure of the distribution(s) at $\beta = 0.3$; this 
is roughly where quantum effects are equal to thermal effects. 
As we can see, for large dissipation ($\Gamma=0.3$) and the most classical 
case, the particle is still confined in one well; that is, thermal 
activation is not high enough to induce switching between wells.
The probability distributions are of course broadened due to thermal
effects, but the signature of the transition as $\beta$ is increased are 
robust. The last row of Fig.~(\ref{grid2}) reassuringly shows 
the existence of a smooth transition away from zero temperature behavior. 

In conclusion, we have shown that in a open double-well quantum oscillator 
clear signatures of quantum behavior and in particular the transition 
away from classical behavior can be found in the changing shapes of 
an asymptotically obtained probability distribution $P_{avg}(x)$. 
At the right parameters, this transition from classical through quantum
mechanical behavior can be nonmonotonic and hence should be clearly 
visible. Further, these transitions are robust at finite temperatures. 
The experimentally tunable parameters in NEMS allow for flexibility in 
exploring the parameter landscape of the transition and are excellent 
candidates to study this transition. We believe that such experiments 
are viable in doubly-clamped NEMS,(see Table.\ref{table1})

{\em Acknowledgements}: The authors gratefully acknowledge funding from
the Howard Hughes Medical Institute through Carleton College, and help 
with computing from Ryan Babbush.

\bibliography{NEMS}
\end{document}